\documentclass{kapproc}
\usepackage{t1enc}
\usepackage{procps} 
\usepackage[dvips]{graphicx}
\upperandlowercase
\kluwerbib

\newcommand{\um}{\,$\mu$m}
\def\gs{\mathrel{\raise0.35ex\hbox{$\scriptstyle >$}\kern-0.6em\lower0.40ex\hbox{{$\scriptstyle \sim$}}}}
\def\ls{\mathrel{\raise0.35ex\hbox{$\scriptstyle <$}\kern-0.6em\lower0.40ex\hbox{{$\scriptstyle \sim$}}}}

\begin{document}

\articletitle{The 2--850\um\ SED of starforming galaxies}

\author{Anna Sajina$^1$, Douglas Scott$^1$, Michel Dennefeld$^2$, Herv\'e Dole$^3$, Mark Lacy$^4$, Guilaine Lagache$^3$}
\affil{$^1$ University of British Columbia, Canada, $^2$ Institut d'Astrophysique de Paris, France, $^3$ Institut d'Astrophysique Spatiale, France, $^4$ Spitzer Science Center, USA}

\begin{abstract}
We present preliminary results on a study of the 2--850\um\ SEDs of a sample of 30 FIRBACK galaxies selected at 170\um. These sources are representative of the brightest $\sim$\,10\% of the Cosmic Infrared Background. They are a mixture of mostly local ($z$\,$<$\,0.3) starforming galaxies, and a tail of ULIGs that extend up to $z$\,$\sim$\,1, and are likely to be a similar population to faint SCUBA sources. We use archival {\sl Spitzer} IRAC and MIPS data to extend the spectral coverage to the mid-IR regime, resulting in an unprecended (for this redshift range) census of their infrared SEDs. This allows us to study in far greater detail this important population linking the near-IR stellar emission with PAH and thermal dust emission. We do this using a Markov Chain Monte Carlo method, which easily allows for the inclusion of \~\,6 free parameters, as well as an estimate of parameter uncertainties and correlations.
\end{abstract}


\section*{Introduction}     
Understanding the full infrared spectral energy distributions (SEDs) of galaxies is essential for a complete picture of star-formation in the Universe. About half the power ever generated by stars is reprocessed by dust into the infrared as inferred from the Cosmic Infrared Background (e.g. Puget et al.~1996; Dwek et al.~1998; Hauser et al.~1998). This emission is increasingly important at higher redshifts where the star-formation density of the Universe is larger than today, coupled with our observational bias toward detecting more active galaxies at high-$z$. 
The variation in SED shapes is a key uncertainty when comparing populations selected at different wavelengths or testing galaxy evolution models. 

We define the infrared SED as covering the range 3--1000\um, which roughly spans the wavelengths between fully stellar and fully synchrotron-dominated emission. Traditionally, studying this entire range at once has been difficult, since mid-IR ($<$\,60\,mJy) observations could not reach much beyond the local Universe (except, to some extend, for the deepest ISOCAM 15\um\ observations), while sub-mm observations also only exist for very local, IR-bright, galaxies (e.g. Dunne et al.~2000), or else SCUBA-selected sources which span $z$\,$\sim$\,1--3 (Chapman et al.~2003; Pope et al.~2005). The latter typically have only one or two detections outside the single SCUBA (850\um) one, which makes their interpretation particularly dependent on the SED model assumed (Blain et al.~2002).  Due to these past observational limitations, we still do not know (beyond some generalized trends) the full range of galaxy SED shapes, how exactly they are related to the underlying physical conditions in the galaxy, and therefore how they may vary across cosmic time as the galaxies evolve.   
        
With the advent of the {\sl Spitzer} Space Telescope \cite{wer04} we can for the first time observe the mid-IR (3\,--\,24\um) properties of large numbers of sources over a cosmologically significant range in $z$ (e.g.~Lonsdale et al.~2003). 
The obvious next step toward characterizing the full infrared SEDs of galaxies is therefore to link {\sl Spitzer} observations with longer wavelength samples, especially including sub-mm observations.

The sample of 30 galaxies discussed here is a sub-sample from the 170\um\ FIRBACK (Far-IR BACKground) ELAIS-N1 catalog (Dole et al.~2001). They were selected on the basis of an existing radio detection which we followed-up with both deep near-IR and sub-mm observations (Scott et al.~2000; Sajina et al.~2003). The sample selection appears to be unbiased with respect to the FIRBACK population as a whole.
Our previous studies suggest that the sample consists primarily of $z$\,$<$\,$0.3$ ordinary spiral-like galaxies rich in cold (T\,$<$\,40\,K) dust, with $\sim$1/6 of the sample consisting of Ultraluminous Infrared Galaxies (ULIGs) at $z$\,$\sim$\,0.5--1. Spectroscopic follow-up of this and related sub-samples confirm both the local sources (Patris et al.~2003; Dennefeld et al.~2004), and the higher-$z$ ones (Chapman et al.~2002). Thus this sample represents a bridge population between the local Universe and SCUBA blank-sky sources (e.g. Pope et al.~2005).          

\begin{figure}
\centering
\vspace*{8.5cm}
\leavevmode
\includegraphics{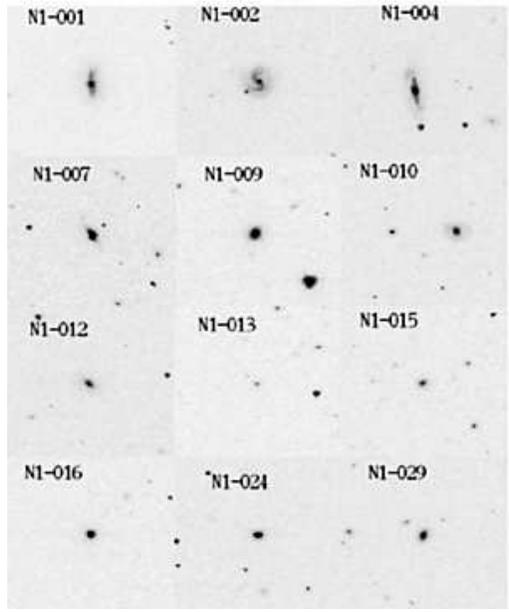}
\caption{\label{mosaic} The brightest 12 of our sources as seen by {\sl Spitzer}, where red, blue, and green are respectively the 8\um, 3.6\um, and 5.8\um\ IRAC channels. The box size is 2$^{\prime}$ (roughly enclosing the {\sl ISO} 170\um\ beam). }
\end{figure}

Here we use archival {\sl Spitzer} observations of the ELAIS-N1 field (see Fig~\ref{mosaic}) in order to extend the known SEDs of the above sample into the mid-IR wavelength range. The IRAC and MIPS plus {\sl ISO} photometry, together with ground-based near-IR and sub-mm data, gives us more than 10 constraints for every galaxy. We fit these SEDs with a phenomenological model motivated by different physical origins for the emission. 
We investigate statistically robust ways to fit the SEDs based on Markov Chain Monte Carlo (MCMC). These allow us to go beyond merely finding the best-fit SED model, toward studying the spread in multi-dimensional, inter-correlated SED models allowed by the data. This not only leads to reliable errors on the derived parameters (e.g. dust temperature, PAH emission strength), but also has obvious implications for the interpretation of data from current and future infrared surveys.   

\section*{SED model \& MCMC fitting}
\begin{figure}
\centering
\vspace*{6.5cm}
\leavevmode
\includegraphics{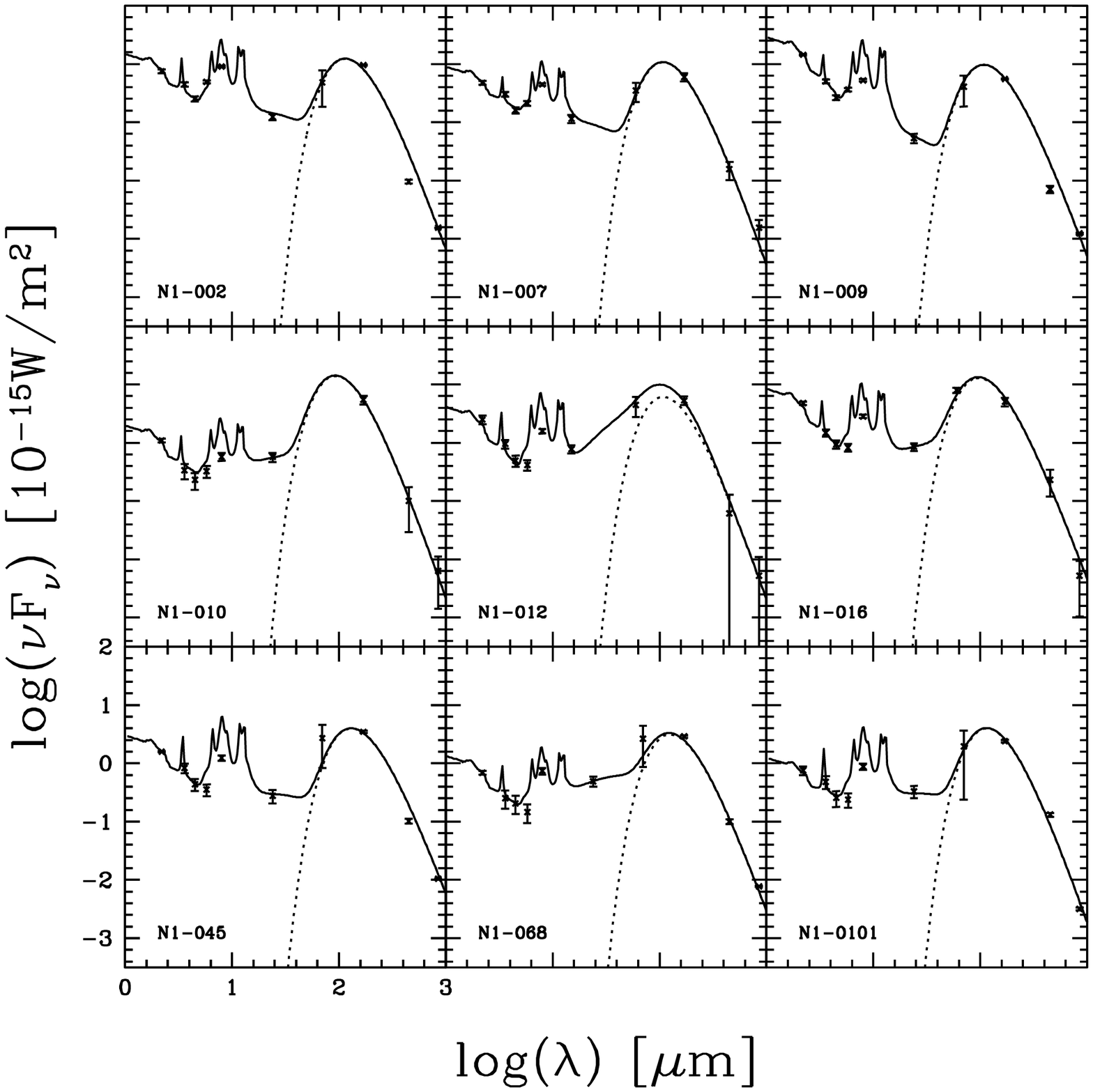}
\includegraphics{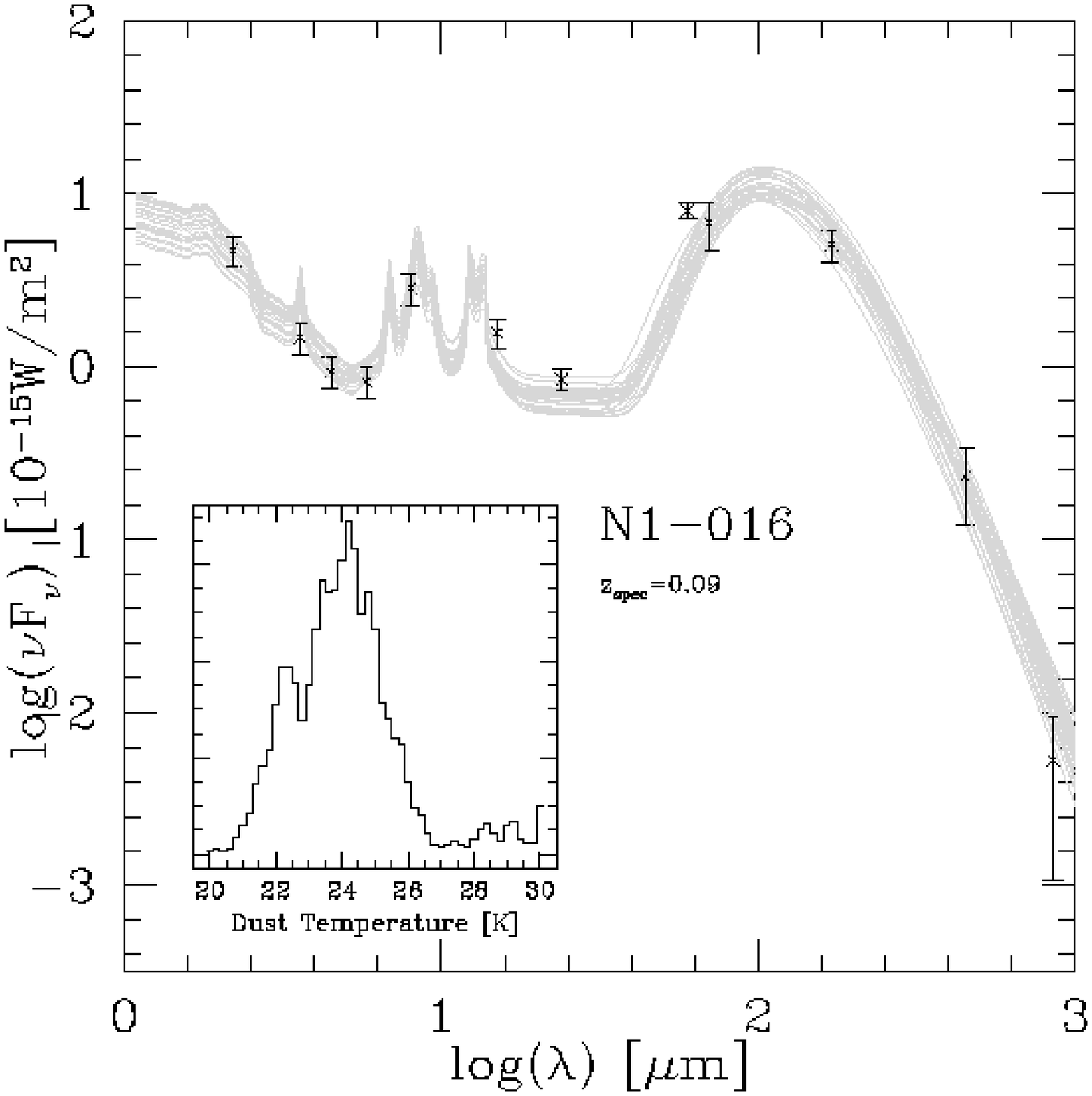}
\caption{\label{seds} Phenomenological modelling of the 2--850\um\ SED of starforming galaxies. On the left, we show a representative sub-sample of our galaxies. The dotted line shows the cold dust component alone. The 8\um\ generally appears lower than the curve, since we account for the filter profiles. On the right, we show an example of the spread in allowable SED shapes from the MCMC fitting. The inset shows the marginalized, posterior probability distribution for the dust temperature.}

\end{figure}

We model the SED as a sum of stellar emission, PAH emission, power-law emission, and thermal grey-body emission. This is similar to the model used in Sajina, Lacy, and Scott~(2005) except for the addition of the thermal component to account for the far-IR and sub-mm emission. The other modification is in the power-law component which now mostly accounts for the 12--100\um\ emission and is cut-off at long-wavelengths in order not to contribute to the far-IR/sub-mm. The stellar emission is accounted for by a 10\,Gyr-old, solar metallicity, Salpeter IMF, single stellar population (SSP) spectral template generated with the PEGASE2.0 spectral synthesis code (Fioc et al.~1997). This amounts to a six parameter model, although a seventh parameter ($\tau_{\rm{V}}$) is needed to fit the ULIG population where optical depths of $\tau_{\rm{V}}$\,$\sim$\,10 are necessary. To find the best-fit model we use the genetic algorithm {\sc pikaia} \cite{pikaia}. Once the loci of interest in the multi-dimensional parameter space have been found, we investigate the probability of the parameters via a Markov Chain Monte Carlo (MCMC) (see Sajina, Lacy, and Scott~2005 and references therein). Fig.~\ref{seds} shows some examples of the model fits and their MCMC scatter.      

\section*{Discussion}   
Here we have presented a new SED model, which aims to be a flexible, template-free approach for investigating the spectral properties of starforming galaxies. The Markov Chain Monte Carlo fitting allows us to sample the posterior probability distribution, providing reliable scatter in the derived parameters as well as revealing any parameter degeneracies. The spectral coverage of the sources shown here is comparable to that expected for upcoming surveys investigating the connections between the mid-IR and far-IR/sub-mm populations (e.g. studies in the GOODS-North and SHADES fields). The advantage of this present sample, however, is that being largely local ($z$\,$<$\,$0.3$), it displays a greater range in intrinsic properties (e.g. SFR), and we already know the redshifts for most of the sources. This also allows us to test various photometric redshift estimators.    
Detailed analysis of the results of this fitting procedure, and implications for the FIRBACK population is work in progress (Sajina et al.~2005b).  

\section*{Acknowledgments}
Most of all we wish to thank the SWIRE team for obtaining, reducing and making publicly available the IRAC and MIPS data we have used in this project. This work is based in part on observations made with the {\sl Spitzer} Space Telescope, which is operated by the Jet Propulsion Laboratory, California Institute of Technology under NASA contract 1407. This research was supported by the Natural Sciences and Engineering Research Council of Canada.

\bibliographystyle{apalike}

\begin{thebibliography}{10}
\bibitem[\protect\citename{Blain et al. }2002]{bl02} Blain A.W., Smail I., Ivison R.J., Kneib J.-P., Frayer D.T.,~2002, PhR, 369, 111 
\bibitem[Chapman et al.(2002)]{scott02} Chapman S.~C., Smail I., Ivison R.~J., Helou G., Dale D.~A., \& Lagache, G.\ 2002, ApJ, 573, 
66
\bibitem[Chapman, Blain, Ivison, \& Smail(2003)]{scott03} Chapman S.~C., Blain A.~W., Ivison R.~J., \& Smail I.~R.\ 2003, Nature, 422, 695 

\bibitem[(Charbonneau et al.~1999)]{pikaia} Charbonneau P.,~1999, ApJ, 101, 309

\bibitem[\protect\citename{Dennefeld et al. }2005]{de05} Dennefeld M., Lagache G., Mei S., et al. 2004, A\&A, submitted

\bibitem[\protect\citename{(Dole et al. }2001)]{d01} Dole H., Gispert R., Lagache G., Puget, J.-L., Bouchet, F. R., Cesarsky, C., Ciliegi, P., Clements, D. L., Dennefeld, M., D\'esert, F.-X.,~2001, A$\&$A, 372, 364

\bibitem[\protect\citename{Dunne, Clements \& Eales }2000]{dce00} Dunne L., Clements D.L., Eales S.A.,~2000, MNRAS, 319, 813
\bibitem[\protect\citename{Dwek et al. }1998]{dw98} Dwek E., Arendt R., Hauser M., Fixsen D., Kelsall T., Leisawitz D., Pei Y. C., Wright E. L., et al.,~1998, ApJ, 508, 106

\bibitem[(Fioc et al.~1997)]{fr97} Fioc M., Roca-Volmerange B.,~1997, A\&A, 326, 950

\bibitem[(Haas et al.~2002)]{haas02} Haas M., Klaas U., \& Bianchi S.,~2002, A\&A, 385, L23
\bibitem[\protect\citename{Hauser et al. }1998]{h98} Hauser M.G., Arendt R.G., Kelsall T., Dwek E., Odegard N., Weiland J. L., Freudenreich H. T., Reach W. T., et al.,~1998, ApJ, 508, 25

\bibitem[\protect\citename{Lagache et al. }2003]{lag02} Lagache G., Dole H., Puget J.-L.,~2003, MNRAS, 338, 555 

\bibitem[(Lonsdale et al.~2003)]{swire} Lonsdale, C.~J., et al.\ 2003, PASP, 115, 897 
\bibitem[(Patris et al.~2003)]{patris} Patris J., Dennefeld M., Lagache G., and Dole H., 2003, A\&A 412, 349
\bibitem[(Pope et al.~2005)]{pope} Pope A., Borys C., Scott D., et al.~2005, MNRAS, in press and these proceedings
\bibitem[(Puget et al.~1996)]{p96} Puget J-L., Abergel A., Bernard J.P., Boulanger F., Burton W. B., D\'esert F.-X., Hartmann D.,~1996, A$\&$A, 308, L5

\bibitem[\protect\citename{Sajina et al. }2003]{me03} Sajina A., Borys C., Chapman S., Dole H., Halpern M., Lagache G., Puget J.-L., Scott D.,~2003, MNRAS, 343, 1365
\bibitem[\protect\citename{Sajina et al. }2005a]{me05} Sajina A., Lacy M., Scott D.,~2005a, ApJ, in press (astro-ph/0409597)
\bibitem[\protect\citename{Sajina et al. }2005]{me05b} Sajina A., Scott D., Dennefeld M. et al.~2005b, in preparation

\bibitem[\protect\citename{Scott et al. }2000]{s00} Scott D., Lagache G., Borys C., Halpern M., Sajina A., Ciliegi P., Clements D. L., et al.,~2000, A$\&$A, 357, L5

\bibitem[(Werner et al.~2004)]{wer04} Werner M., et al.~2004, ApJS

\end{thebibliography}
\chapbblname{sajinaa}
\chapbibliography{logic}

\end{document}